\documentclass[aps,preprint,showpacs,nofootinbib,floatfix,superscriptaddress,preprintnumbers]{revtex4}
\usepackage{amsmath}
\usepackage{amssymb}
\usepackage{epsfig}
\usepackage{graphicx}
\usepackage{stmaryrd}
\usepackage{psfrag}
\usepackage{setspace}
\usepackage{color}

\newcommand {\black} {\color{black}}

\def\lsim{\raise0.3ex\hbox{$\;<$\kern-0.75em\raise-1.1ex\hbox{$\sim\;$}}}
\def\vb#1{\vbox to #1 pt{}}
\def\vev#1{\left\langle #1\right\rangle}
\def\321{SU(3) $\otimes$ SU(2) $\otimes$ U(1)}
\newcommand{\be}{\begin{equation}}
\newcommand{\ee}{\end{equation}}
\newcommand{\bd}{\begin{displaymath}}
\newcommand{\ed}{\end{displaymath}}
\newcommand{\bea}{\begin{eqnarray}}
\newcommand{\eea}{\end{eqnarray}}

\newcommand {\ignore}[1]{}

\newcommand{\AddrAHEP}{%
  AHEP Group, Institut de F\'{\i}sica Corpuscular --
  C.S.I.C./Universitat de Val{\`e}ncia \\
  Edificio Institutos de Paterna, Apt 22085, E--46071 Valencia, Spain}
\newcommand{\AddrLisb}{%
 Departamento de F\'\i sica and CFTP, Instituto Superior T\'ecnico\\
          Av. Rovisco Pais 1, 1049-001 Lisboa, Portugal }
\newcommand{\AddrCERN}{CERN, Theory Division, CH-1211 Geneva 23, Switzerland}

\begin{document}\setlength{\unitlength}{1mm}

\preprint{CERN-PH-TH/2010-132} \preprint{CFTP/10-008}
\preprint{IFIC/10-18}

\title{ $A_4$-based neutrino masses with Majoron decaying dark
  matter} \date{\today}

\author{J.~N.~Esteves}\email{joaomest@cftp.ist.utl.pt}\affiliation{\AddrLisb}
\author{F.~R.~Joaquim}\thanks{On leave from ``Centro de F\'{i}sica Te\'orica de Part\'{i}culas (CFTP)'', Lisbon, Portugal.}\email{Filipe.Joaquim@cern.ch}\affiliation{\AddrCERN}
\author{A.~S.~Joshipura} \email{anjan@prl.ernet.in} \affiliation{Physical Research Laboratory, Ahmedabad 380009, India}
\author{J.~C.~Rom\~ao} \email{jorge.romao@ist.utl.pt}\affiliation{\AddrLisb}
\author{M.~A.~T\'ortola}\email{mariam@ific.uv.es}\affiliation{\AddrAHEP}
\author{J.~W.~F.~Valle} \email{valle@ific.uv.es} \affiliation{\AddrAHEP}


\begin{abstract}
  We propose an $A_4$ flavor-symmetric \321 seesaw model where lepton
  number is broken spontaneously.  A consistent two-zero texture
  pattern of neutrino masses and mixing emerges from the interplay of
  type-I and type-II seesaw contributions, with important
  phenomenological predictions.  We show that, if the Majoron becomes
  massive, such seesaw scenario provides a viable candidate for
  decaying dark matter, consistent with cosmic microwave background
  lifetime constraints that follow from current WMAP observations.  We
  also calculate the sub-leading one-loop-induced decay into photons
  which leads to a mono-energetic emission line that may be observed
  in future X-ray missions such as Xenia.

\end{abstract}

\maketitle

\section{Introduction}

The discovery of neutrino
oscillations~\cite{fukuda:2002pe,ahmad:2002jz,eguchi:2002dm,fukuda:1998mi,Kajita:2004ga},
now confirmed at reactors and
accelerators~\cite{Ahn:2006zza,Adamson:2008zt,abe:2008ee}, has
brought neutrino physics to the center of particle physics research.
Global analysis of current oscillation data indicate that the
pattern of lepton mixing differs sharply from that characterizing
quarks~\cite{Schwetz:2008er}.  Understanding the origin of neutrino
mass and the pattern of neutrino mixing angles from basic principles
constitutes a major
challenge~\cite{Altarelli:2004za,Nunokawa:2007qh}. A paradigm
framework to generate neutrino masses is provided by the seesaw
mechanism, for which several realizations have been
proposed~\cite{Valle:2006vb}. The observed pattern of neutrino
mixing may arise from suitable non-abelian flavour symmetries, as
those based on the $A_4$
group~\cite{Ma:2001dn,babu:2002dz,Zee:2005ut,Altarelli:2005yx}.

Elucidating the nature of dark matter constitutes another intriguing
problem of modern physics which has so far defied all efforts. It is
therefore crucial to build a fundamental particle physics theory of
dark matter and,
since the Standard Model of elementary particles (SM) fails to
provide a dark matter candidate, such theory necessarily requires
physics beyond the SM.

Here we suggest a version of the seesaw mechanism containing both
type-I~\cite{Minkowski:1977sc,GellMann:1980vs,Yanagida:1979as,Glashow:1979nm,
  Mohapatra:1979ia,Schechter:1980gr,Schechter:1981cv,chikashige:1981ui}
and type-II
contributions~\cite{Schechter:1980gr,Schechter:1981cv,cheng:1980qt,Magg:1980ut,Lazarides:1980nt,
  Mohapatra:1980yp} in which we implement an $A_4$ flavor symmetry
with spontaneous violation of lepton
number~\cite{chikashige:1981ui,Schechter:1981cv}. We study the
resulting pattern of vacuum expectation values (vevs) and show that
the model reproduces the phenomenologically consistent and
predictive two-zero texture proposed in Ref.~\cite{Hirsch:2007kh}.

In the presence of explicit global symmetry breaking effects, as
might follow from gravitational interactions, the resulting
pseudo-Goldstone boson - Majoron - may constitute a viable candidate
for decaying dark matter if it acquires mass in the keV-MeV range.
Indeed, this is not in conflict with the lifetime constraints which
follow from current cosmic microwave background (CMB) observations
provided by the Wilkinson Microwave Anisotropy Probe
(WMAP)~\cite{Komatsu:2008hk}.
We also show how the corresponding mono-energetic emission line
arising from the sub-leading one-loop induced electromagnetic decay
of the Majoron may be observed in future X-ray
missions~\cite{herder:2009im}.

The paper is organized as follows. In section \ref{sec:A4model} we
describe our $A_4$ model while in section \ref{sec:A4pot} we discuss
the symmetry breaking structure which is required to obtain the
correct neutrino texture. In section \ref{sec:neutrino}, we update
the neutrino parameter analysis and we study the implications of a
decaying Majoron dark matter scenario in section
\ref{sec:MajoronDM}. Further discussion is presented in the
concluding section~\ref{sec:conclusions}.

\section{The Model}
\label{sec:A4model}

Our model is described by the multiplet content specified in Table
\ref{tab:QuantumNumbers} where the transformation properties under
the SM and $A_4$ groups are shown (as well as the corresponding
lepton number $L$).  The $L_i$ and $l_{Ri}$ fields are the usual SM
lepton doublets and singlets and $\nu_R$ the right-handed neutrinos.
The scalar sector contains an SU(2) triplet $\Delta$, three Higgs
doublets $\Phi_i$ (which transform as a triplet of $A_4$) and a
scalar singlet $\sigma$. Three additional fermion singlets $S_i$ are
also included.
\begin{table}[ht!]
  \centering
  \caption{Lepton multiplet structure ($Q=T_3+Y/2$)}
  \begin{tabular}{ccccccccccc}\hline\hline
 \hskip 10mm
&\phantom{\hskip 2mm}$L_1$\phantom{\hskip 2mm} &\phantom{\hskip
2mm}$L_2$\phantom{\hskip 2mm} &\phantom{\hskip
2mm}$L_3$\phantom{\hskip 2mm} &\phantom{\hskip
2mm}$l_{Ri}$\phantom{\hskip 2mm} &\phantom{\hskip
2mm}$\nu_{iR}$\phantom{\hskip 2mm} &\phantom{\hskip
2mm}$\Phi_i$\phantom{\hskip 2mm} &\phantom{\hskip
2mm}$\Delta$\phantom{\hskip 2mm}
&\phantom{\hskip 2mm}$\sigma$\phantom{\hskip 2mm}
&\phantom{\hskip 2mm}$S_i$\phantom{\hskip 2mm}\\[+2pt] \hline
$SU(2)$&$2$&$2$&$2$&$1$&$1$&$2$&$3$ &$1$ &$1$ \\[+2pt]
$U(1)_Y$&$-1$&$-1$&$-1$&$-2$&$0$&$-1$&$2$ &$0$ &$0$ \\[+2pt]
\hline
$A_4$&$1'$&$1$&$1''$&$3$&$3$&$3$&$1''$ & $1''$ & $3 $\\[+2pt]
\hline
$L$&1&$1$&$1$&$1$&$1$&$0$&$-2$ & $-2$ & $1$\\[+2pt]\hline\hline
  \end{tabular}
  \label{tab:QuantumNumbers}
\end{table}

Taking into account the information displayed in
Table~\ref{tab:QuantumNumbers}, and imposing lepton number
conservation, the Lagrangian responsible for neutrino masses reads
\begin{align}
  \label{eq:Lagneutrmasses}
 -\mathcal{L}_{L}&= h_{1} \overline{L}_1 \left(\nu_R
    \Phi\right)'_1
+h_{2} \overline{L}_2 \left(\nu_R \Phi\right)_1
+h_{3} \overline{L}_3 \left(\nu_R \Phi\right)''_1\nonumber \\
&+\lambda L_{1}^{T}C\Delta L_{2}+ \lambda L_{2}^{T}C\Delta L_{1} +
\lambda^{\prime}
L_{3}^{T}C\Delta L_{3}\nonumber\\
&+ M_R \left(\overline{S_L} \nu_R\right)_1 + h \left(S^T_L C
S_L\right)'_1\sigma+\hbox{h.c.}\,,
\end{align}
where $h$ and $\lambda$ are adimensional couplings, $M_R$ is a mass
scale and
\begin{align}
\label{eq:PhiDelta}
    \Delta=\left(
  \begin{array}{cc}
    \Delta_0 & -\Delta^+/\sqrt{2}  \\
    -\Delta^+/\sqrt{2} &\Delta^{++}
  \end{array}
\right)\,,\,\Phi_i=\left(
  \begin{array}{c}
    \phi_i^0  \\
    \phi_i^{-}
  \end{array}
\right)\,.
\end{align}
Note that the term ($\nu_R^T C \nu_R)'_1\sigma$ is
 allowed by the imposed symmetry. This term however does not
   contribute to the light neutrino masses to the leading order in the
  seesaw expansion and we omit it.
  Alternatively, such term may be forbidden by holomorphy in a
  supersymmetric framework with the following superpotential terms
\begin{equation}\label{eq:SusyPot}
    \mathcal{W}=\dots+\lambda
    \epsilon_{ab}h_i^\nu\hat{L}^a_i\hat{\nu}^c\hat{H}_u^b+ M_R \hat{\nu}^c\hat{S}
    +\frac{1}{2} h \hat{S}\hat{S}\hat{\sigma}  \nonumber
\end{equation}
where the hats denote superfields and the last term replaces the
corresponding bilinear employed in
Ref.~\cite{mohapatra:1986bd,gonzalez-garcia:1989rw}.
Assuming that the Higgs bosons $\Phi_i$, $\Delta^0$ and $\sigma$
acquire the following vevs (see section \ref{sec:A4pot} below)
\begin{equation}
\label{eq:2} \vev{\phi^0_1}=  \vev{\phi^0_2}=  \vev{\phi^0_3}=
\frac{v}{\sqrt{3}}, \quad \vev{\Delta^0} = u_{\Delta}, \quad
\vev{\sigma} = u_{\sigma}\,,
\end{equation}
we obtain an extended seesaw neutrino mass matrix
$\mathcal{M}$~\cite{mohapatra:1986bd,gonzalez-garcia:1989rw,deppisch:2004fa}
in the ($\nu_L$, $\nu^c$, S) basis
\begin{equation}\label{eq:neutrmassmatrix}
\mathcal{M}= \left(\begin{array}{ccc}
0 & m_D & 0 \\
m_D^T & 0 & M \\
0 & M^T & \mu \\
\end{array}\right)\;,\;m_D= v\ \hbox{diag}(h_{1},h_{2},h_{3})\ U, \quad
U=\frac{1}{\sqrt{3}}\left(
\begin{array}{ccc}
1 &\omega^2 & \omega\\[+2mm]
1  & 1 & 1\\[+2mm]
1 & \omega &\omega^2
\end{array}
\right)\,,
\end{equation}
with $\omega=e^{2\pi i/3}$, $M=M_R\hspace{0.1cm}\text{diag}(1,1,1)$
and $\mu=u_\sigma h\hspace{0.1cm} \text{diag}(1,w^2,w)$.
This leads to an effective light neutrino mass matrix
$\mathcal{M}_\nu^{\rm I}$ given by
\begin{equation}
\label{eq:invseesaw} \mathcal{M}^{\rm I}_{\nu}=m_D{M^T}^{-1}\mu
M^{-1}m_D^T=
\frac{h v^2 u_{\sigma}}{M_R^2} \left(
\begin{array}{ccc}
 h_{1}^2 & 0 & 0 \\[+2mm]
0 & 0 & h_{2} h_{3} \\[+2mm]
0 & h_{2} h_{3}& 0
\end{array}
\right)\,.
\end{equation}
On the other hand the vev of the triplet, $u_{\Delta}$,  will induce
an effective mass matrix for the light neutrinos from type-II seesaw
mechanism
\begin{equation}
\label{eq:II}
 \mathcal{M}^{\rm II}_{\nu}=2u_{\Delta}\left(
    \begin{array}{ccc}
      0 &\lambda  & 0\\[+2mm]
      \lambda & 0 & 0\\[+2mm]
      0 & 0 & \lambda^{\prime}
    \end{array}
\right)\,,
\end{equation}
and the total effective light neutrino mass matrix will then be
\begin{equation}
  \label{eq:T}
  \mathcal{M}_{\nu}=\mathcal{M}^{\rm I}_{\nu}+\mathcal{M}^{\rm
    II}_{\nu}\,.
\end{equation}

In Ref.\cite{Hirsch:2007kh} it was shown that the neutrino mass
matrix given by Eq.~(\ref{eq:T}) could explain the currently
available neutrino data. In section~\ref{sec:neutrino} we will
present an update of that analysis taking into account the latest
neutrino oscillation data.

\section{$A_4$ Invariant Higgs Potential}
\label{sec:A4pot}

We now address the question of the minimization of the neutral Higgs
scalar potential, which is a necessary condition to reproduce the
structure of the neutrino mass matrix presented in the previous
section.  With the assignments of Table~\ref{tab:QuantumNumbers},
the Higgs potential consistent with gauge and $A_4$ invariance and
lepton number conservation reads,
\begin{equation}
\label{eq:HiggsPot} V=V(\Phi)+V(\Phi,\Delta,\sigma)\,,
\end{equation}
where $V(\Phi)$ is given as (the decomposition of the tensorial
product of two triplets in $A_4$ is shown in the Appendix):
\begin{align}
\label{eq:VPhi} &\notag V(\Phi)=m_\Phi^2\left(\Phi^\dag\Phi\right)_1
+\lambda_1\left(\Phi^\dag\Phi\right)_1\left(\Phi^\dag\Phi\right)_1
+\lambda_2\left(\Phi^\dag\Phi\right)_{1'}\left(\Phi^\dag\Phi\right)_{1''}\\
&+\lambda_3\left(\Phi^\dag\Phi\right)_{3s}\cdot\left(\Phi^\dag\Phi\right)_{3s}
+\lambda_4\left(\Phi^\dag\Phi\right)_{3s}\cdot\left(\Phi^\dag\Phi\right)_{3a}
+\lambda_5\left(\Phi^\dag\Phi\right)_{3a}\cdot\left(\Phi^\dag\Phi\right)_{3a}\,,
\end{align}
and $V(\Phi,\Delta,\sigma)$ contains pure $\Delta\,,\sigma$ terms,
together with others involving mixed invariant combinations of the
scalar fields.
Assuming the so-called seesaw hierarchy $u_\Delta\ll v \ll
u_\sigma$~\cite{Schechter:1981cv}~\footnote{In contrast to the
inverse
  seesaw models used in
  Refs.~\cite{gonzalez-garcia:1989rw,deppisch:2004fa} here we consider
  large values of $u_\sigma$, $u_\sigma > 10^7$~GeV or so.}, the
relevant terms in $V(\Phi,\Delta,\sigma)$ are~\footnote{Notice that
  the scalar potential contains other invariant terms such as
  $\Phi^\dag \Phi{\rm Tr}(\Delta^\dag\Delta)$, ${\rm
    Tr}(\Delta^\dag\Delta)|\sigma|^2$, $[{\rm
    Tr}(\Delta^\dag\Delta)]^2$, etc.  Assuming the vev hierarchy
  $u_\Delta<<v<<u_\sigma$ and that the adimensional coefficients of
  these terms are of the same order of the ones in
  $V(\Phi,\Delta,\sigma)$, then $V(\Phi,\Delta,\sigma)$ is enough for
  our purposes. }
\begin{equation}
\label{eq:VPhiTphi}
 V(\Phi,\Delta,\sigma)=\left(M_\Delta^2+\rho|\sigma|^2\right)\text{Tr}(\Delta^\dag
\Delta)
+\lambda_{\sigma}|\sigma|^4+\left[m_{\sigma}^2+\xi\left(\Phi^\dag\Phi\right)_{1}
\right]|\sigma|^2-(\delta\Phi^T \Delta\Phi\sigma^* + \text{h.c.}),
\end{equation}
Taking the vacuum alignment for the Higgs doublets $\Phi_a$ given in
eq.~(\ref{eq:2}) the minimization of the Higgs potential with
respect to $\Delta$ gives
\begin{equation}
\label{eq:dervTaprox} \frac{\delta V}{\delta \Delta}=0 \Rightarrow
(M_\Delta^2+\rho \,u_\sigma^2)\,u_\Delta - {\delta} v^2 u_\sigma
=0\,.
\end{equation}
We stress that the $A_4$ symmetry, together with the doublet vev
alignment assumed in Eq.~(\ref{eq:2}), requires that the product
$\Phi\otimes\Phi\sim{\bf 1}$ under $A_4$. If
$\Phi\otimes\Phi\sim{\bf
  1^\prime,\,1^{\prime\prime}}$, then the second term in the above
equation would reduce to $2{\delta} (1+\omega+\omega^2) u_\sigma=0$
implying $u_\Delta\sim 0$.  Moreover, as a direct consequence of the
requirement $\Phi\otimes\Phi\sim{\bf 1}$ under $A_4$, $\Delta$ and
$\sigma$ must have the same (singlet) transformation properties
under that group.

The above equation leads to the following solution for the triplet
vev
\begin{equation}
\label{eq:vevT} u_\Delta=\frac{{\delta} v^2
u_\sigma}{M_\Delta^2+\rho u_\sigma^2}\simeq \frac{{\delta} v^2}{\rho
u_\sigma}\,,
\end{equation}
where the last approximation holds for $M_\Delta\ll u_\sigma$. This
result shows that the ``vev-seesaw'' relation $u_\Delta u_\sigma\sim
v^2$ is fulfilled.
The minimization with respect to the $\Phi_a$ gives
\begin{equation}
  \label{eq:derVPhi}
  \frac{\delta V}{\delta \Phi_a}=0 \Rightarrow \frac{\delta V(\Phi)}{\delta
    \Phi_a}+2\xi v u_\sigma^2 -4{\delta} v u_\Delta u_\sigma =0.
\end{equation}

Finally,
\begin{equation}
\label{eq:derVphi} \frac{\delta V}{\delta \sigma}=0 \Rightarrow
2\lambda_\sigma u_\sigma^3 + \left(m_\sigma^2 +\,\xi v^2+\rho
u_\Delta^2\right)u_\sigma -2{\delta} v^2 u_\Delta =0.
\end{equation}
which, in the limit $u_\Delta,v<<u_\sigma$, has the approximate
solution
\begin{equation}
\label{eq:vevphi}
u_\sigma=\sqrt{-\frac{m_\sigma^2}{2\lambda_\sigma}},
\end{equation}
as it is typical from spontaneous symmetry breaking scenarios. In
summary, we have shown that in our framework it is possible to
achieve a consistent minimization of the scalar potential with
non-zero vevs satisfying the ``vev-seesaw'' relation $u_\Delta
u_\sigma\sim v^2$.

\section{Neutrino parameter analysis}
\label{sec:neutrino}

Given the two contributions to the light neutrino mass matrix
discussed in Eqs.~(\ref{eq:invseesaw}) and (\ref{eq:II}) one finds
that the total neutrino mass matrix has the following structure:
\begin{equation}
 \mathcal{M}_{\nu}=\left(
    \begin{array}{ccc}
      a & b & 0\\[+2mm]
      b & 0 & c\\[+2mm]
      0 & c & d
    \end{array}
\right). \label{eq:B1}
\end{equation}
This matrix with two-zero texture has been classified as B1 in
\cite{Frampton:2002yf}. One can show that considering the
$(L_1,L_2,L_3)$ transformation properties under $A_4$ as being
$(1^\prime,1^{\prime\prime},1)$ or $(1^{\prime\prime},1^{\prime},1)$
an effective neutrino mass matrix with
$\mathcal{M}_{\nu}(1,2)=\mathcal{M}_{\nu}(3,3)=0$ is obtained (type
B2 in~\cite{Frampton:2002yf}). Moreover, by choosing
$\Delta,\sigma\sim {\bf 1^\prime}$ and appropriate transformation
properties of the $L_i$ doublets, we could obtain the textures B1
and B2 as well. Still, the configuration $\Delta,\sigma\sim {\bf 1}$
would lead to textures which are incompatible with neutrino data
since, in this case, both type I and type II contributions to the
effective neutrino mass matrix would have the same form. Since the
textures of the type B1 and B2 are very similar in what concerns to
neutrino parameter predictions, we will restrict our analysis to B1,
shown in (\ref{eq:B1}).

In general, the neutrino mass matrix is described by nine
parameters: three masses, three mixing angles and three phases (one
Dirac + two Majorana). From neutrino oscillation experiments we have
good determinations for two of the mass parameters (mass squared
differences) and for two of the mixing angles ($\theta_{12}$ and
$\theta_{23}$) as well as an upper-bound on the third mixing angle
$\theta_{13}$.  Using the 3$\sigma$ allowed ranges for these five
parameters and the structure of the mass matrix in Eq.~(\ref{eq:B1})
we can determine the remaining four parameters.
The phenomenological implications of this kind of mass matrix have
been analysed in Refs. \cite{Hirsch:2007kh} and \cite{Dev:2006qe}.
Here we will update the results in light of the recently determined
neutrino oscillation parameters~\cite{Schwetz:2008er}.

\begin{figure}[t]
  \centering
  \begin{tabular}{ccc}
    \includegraphics[height=5.5cm]{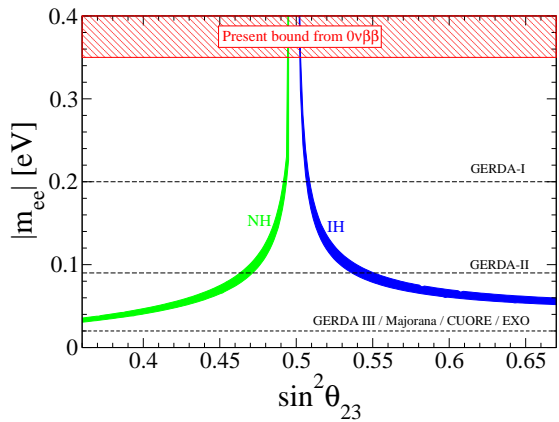}&
  \end{tabular}
  \caption{Correlation between the neutrinoless double beta decay
    amplitude parameter $|m_{ee}|$ and the atmospheric mixing
    parameter.  Experimental sensitivities are also given for
    comparison.}
  \label{fig:texture1}
\end{figure}

\begin{figure}[!h]
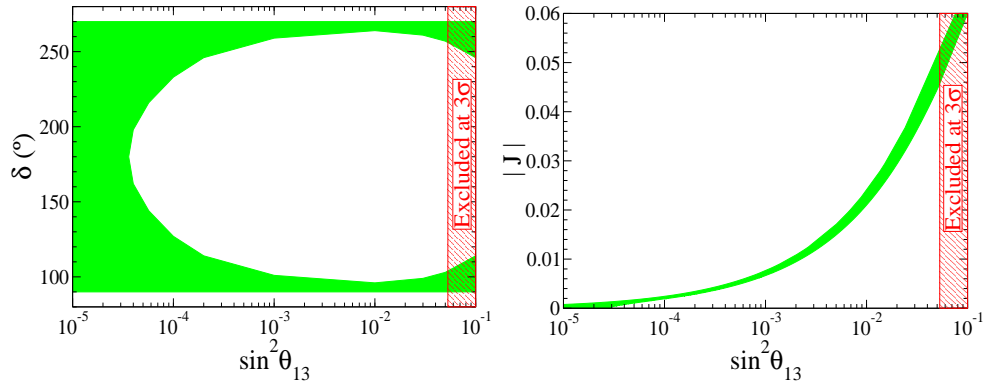

  \centering
  \begin{tabular}{ccc}
    \includegraphics[height=5cm,width=0.4\linewidth]{Fig1b-solid.eps}&
    \includegraphics[height=5cm,width=0.4\linewidth]{Fig1c-solid.eps}
  \end{tabular}
  \caption{CP violating phase $\delta$ and CP-invariant $J$ in terms
    of the reactor mixing parameter.  The 3~$\sigma$-excluded range
    for  $\sin^2\theta_{ij}$ is given for comparison.}
  \label{fig:texture2}
\end{figure}

The main results are shown in Figs.~\ref{fig:texture1} and
\ref{fig:texture2}.  In figure~\ref{fig:texture1} we plot the
correlation of the mass parameter characterizing the neutrinoless
double beta decay amplitude:
\begin{equation}
\left| m_{ee} \right| = \left| c^2_{13} c^2_{12} m_1 +  c^2_{13}
  s^2_{12} m_2 e^{2i\alpha} + s^2_{13} m_3 e^{2i\beta} \right|,
\end{equation}
with the atmospheric mixing angle $\theta_{23}$.  Here $c_{ij}$ and
$s_{ij}$ stand for $\cos\theta_{ij}$ and $\sin\theta_{ij}$
respectively.  At the zeroth order approximation $m_1/m_3 =
\tan^2\theta_{23}$, and therefore $\theta_{23} < 45^\circ$ for
normal hierarchy (NH), while $\theta_{23} > 45^\circ$ for inverted
hierarchy (IH).  The main result from this plot is a lower bound on
the effective neutrino mass:$\left| m_{ee} \right| > 0.03$ eV.  For
comparison the range of sensitivities of planned experiments as well
as current bounds is also given.
Note that the lower bound we obtain lies within reach of the future
generation of neutrinoless double beta decay experiments.

The panels in Fig.~\ref{fig:texture2} show the CP-violating phase
$\delta$ and the corresponding CP-violating invariant in neutrino
oscillations:
 \begin{equation}
J = s_{12} s_{23} s_{13} c_{12} c_{23} c^2_{13} \sin\delta\,,
\end{equation}
versus $\sin^2\theta_{13}$.
Note that these hold both for normal and inverted hierarchy spectra.
In the middle panel one sees that $\cos\delta < 0$ since, at first
order in $\sin^2\theta_{13}$, $m_1/m_2 = 1 +
\frac{\cos\theta_{23}}{\cos\theta_{12}\sin\theta_{12}\sin^2\theta_{23}}\sin\theta_{13}\cos\delta$,
and the ratio of masses should satisfy: $m_1/m_2 < 1$. Moreover, for
large $\theta_{13}$ values, where CP violation is likely to be
probed in neutrino oscillations, one can see that our model predicts
maximal violation of CP. Quantitatively, from the right panel one
sees that the 3$\sigma$ bound on $\theta_{13}$: $\sin^2\theta_{13} <
0.053$ implies an upper bound: $|J| \lesssim 0.06$ on the
CP-invariant.

In addition, the two-zero texture structure of our neutrino mass
matrix may have other implications, for example for the expected
pattern of lepton flavor violating decays. In fact, thanks to the
strong renormalization effects due to the presence of the triplet
states, the latter are quite sizeable in sypersymmetric
models~\cite{Rossi:2002zb,Joaquim:2006uz, Esteves:2009vg}.

\section{Majoron Dark Matter}
\label{sec:MajoronDM}

In models where neutrinos acquire mass through spontaneous breaking
of an ungauged lepton
number~\cite{chikashige:1981ui,Schechter:1981cv} one expects that,
due to non-perturbative effects, the Nambu-Goldstone boson (Majoron)
may pick up a mass that we assume to lie in the kilovolt
range~\cite{berezinsky:1993fm}. This implies that the Majorons will
decay, mainly in neutrinos. As the coupling $g_{J\nu\nu}$ is
proportional to  $\frac{m_\nu}{u_\sigma}$~\cite{Schechter:1981cv},
the corresponding mean lifetime can be extremely long, even longer
than the age of the Universe. As a result the Majoron can, in
principle, account for the observed cosmological dark matter (DM).

This possibility was explored in
Refs.~\cite{lattanzi:2007ux,bazzocchi:2008fh} in a general context.
Here, we just summarize the results. It was found that the relic
Majorons can account for the observed cosmological dark matter
abundance provided
\begin{equation}
  \label{eq:7}
  \Gamma_{J\nu\nu} < 1.3 \times 10^{-19}\ \hbox{s}^{-1}\,\,,\,\,
0.12\ \hbox{keV} < \beta \,m_J < 0.17\ \hbox{keV}\,,
\end{equation}
where $\Gamma_{J\nu\nu}$ is the decay width of $J\rightarrow \nu\nu$
and $m_J$ is the Majoron mass. The parameter $\beta$ encodes our
ignorance about the number density of Majorons, being normalized to
$\beta=1$ if the Majoron was in thermal equilibrium in the early
Universe decoupling sufficiently early, when all other degrees of
freedom of the standard model were excited~\cite{bazzocchi:2008fh}.
In the following we will follow their choice and will take
\begin{equation}
  \label{eq:9}
  10^{-5} < \beta < 1,
\end{equation}
and calculate both the width into neutrinos as well as the
subleading one-loop induced decay into photons.

\subsection{Decay into neutrinos}

We now proceed with the computation of the Majoron decay width into
neutrinos, which will be useful to obtain the allowed parameter
space for which the Majoron can be a viable DM candidate. In order
to calculate the decay amplitude we remind that the coupling
$g_{J\nu_i\nu_j}$ is defined through
\begin{equation}
  \label{eq:10}
  \mathcal{L} = - \frac{1}{2}g_{J\nu_i\nu_j} J \nu_i \nu_j  + \hbox{\ h.c.}
\end{equation}
For the evaluation of $g_{J\nu_i\nu_j}$, we follow the steps
developed in Ref.~\cite{Schechter:1981cv}. First we notice that with
scalar potential defined in section~\ref{sec:A4pot}, the Majoron, in
the basis $\left[\mathrm{Im}(\phi_i^0),\mathrm{Im}(\Delta^0),
  \mathrm{Im}(\sigma^0)\right]^T$, is given by
\begin{equation}
  \label{eq:11}
  J = N_J \left[ 2 u_{\Delta}^2 \frac{v}{\sqrt{3}},
  2 u_{\Delta}^2 \frac{v}{\sqrt{3}},
  2 u_{\Delta}^2 \frac{v}{\sqrt{3}},
   u_{\Delta} v^2, u_{\sigma} (4 u_{\Delta}^2 + v^2)\right]\,,
\end{equation}
and
\begin{equation}
  \label{eq:12}
  N_J= \left[4 v^2 u_{\Delta}^4 +  v^4 u_{\Delta}^2 + u_{\sigma}^2 (4
    u_{\Delta}^2 + v^2)^2\right]^{-1/2} \simeq \frac{1}{v^2 u_{\sigma}}\,,
\end{equation}
where the last equality follows from the assumed hierarchy
$u_{\Delta} \ll v \ll u_{\sigma}$ implied by the vev-seesaw
relation. Using this, one can obtain
\begin{equation}
  \label{eq:41}
  g_{J\nu_i\nu_j} = - \frac{m^\nu_i \delta_{ij}}{\sqrt{2}\ u_{\sigma}}\,,
\end{equation}
leading to the decay width
\begin{equation}
  \label{eq:42}
  \Gamma_{J\nu\nu} = \frac{m_J}{32\pi} \frac{\sum_i (m^\nu_i)^2}{2
    u_{\sigma}^2} \ .
\end{equation}
It is worth mentioning that the sum $\sum_i (m^\nu_i)^2$ is in our
framework constrained by the special form of the effective neutrino
mass matrix shown in Eq.~(\ref{eq:B1}). In particular, there is a
lower bound on the mass of the lightest neutrino: $m \gtrsim 0.03$
eV.

\subsection{Decay into photons}
The Majoron also couples with photons (at the quantum level) and
therefore the radiative decay $J\rightarrow \gamma \gamma$ is
expected to occur with a photon energy $E_\gamma \simeq m_J/2$.
Consequently, this decay exhibits a mono-energetic emission line
which could be detected in a variety of X-ray observatories, see for
example the discussion given in
Refs.~\cite{herder:2009im,bazzocchi:2008fh}.

The effective Majoron-photon interaction can be written as
\begin{figure}[!htb]
  \centering
\includegraphics[width=0.65\textwidth]{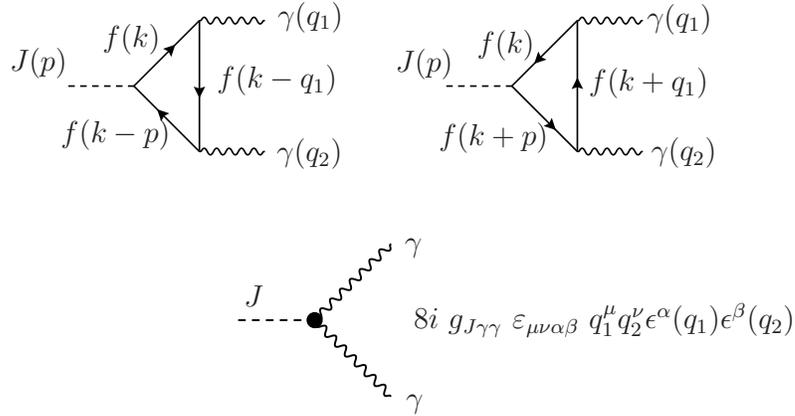}
\caption{Top: One loop diagrams for the decay $J\rightarrow
\gamma\gamma$. Bottom: Effective $J\gamma\gamma$ vertex.}
\label{fig:oneloop}
\end{figure}
\begin{equation}
  \label{eq:1}
  \mathcal{L} = g_{J\gamma\gamma} \varepsilon^{\mu\nu\alpha\beta}
  F_{\mu\nu} F_{\alpha\beta}\,,
\end{equation}
resulting from the one-loop diagrams shown in Fig.~\ref{fig:oneloop}
(top diagrams). The effective coupling $g_{J\gamma\gamma}$ (bottom
graph in Fig.~\ref{fig:oneloop}) is

\begin{equation}
  \label{eq:4}
  g_{J\gamma\gamma}^f \equiv \frac{N_f \alpha^2 g_{J f f} Q_f^2 X_f}{8\pi m_f}\,,
\end{equation}
with $ X_f = -2 m_f^2 C_0(0, 0, m_J^2 , m_f^2 , m_f^2 , m_f^2)\simeq
1 + m_J^2/(12 m_f^2)$ where $C_0$ is the invariant Passarino-Veltman
loop function~\cite{passarino:1979jh}.  The last approximation is
valid for $m_J \ll m_f$. $T_3^f$, $Q_f$ and $N_f$ denote the weak
isospin, the electric charge and the colour factor of the
corresponding charged fermion $f$, respectively.  The coupling of
the Majoron to the charged fermions $g_{Jff}$ is given
by~\cite{bazzocchi:2008fh}
\begin{equation}
  \label{eq:5}
  g_{J f f} = - \frac{2 u_{\Delta}^2}{v^2 u_{\sigma}} m_f (- 2 T_{3}^f)\,.
\end{equation}
We then get for the decay width,
\begin{eqnarray}\black
  \nonumber
  \Gamma_{J\gamma\gamma} & = & \frac{m_J^3}{\pi}
  \left| \sum_f g_{J\gamma\gamma}^f \right|^2 =
\frac{\alpha^2 m_J^3}{64\pi^3} \left| \sum_f  \frac{ N_f g_{J f f}
Q_f^2
    X_f}{ m_f}\right|^2 = \\
 & = & \frac{\alpha^2 m_J^3}{64\pi^3}  \left| \sum_f N_f  Q_f^2
\frac{2 u_{\Delta}^2}{v^2 u_{\sigma}} (- 2 T_{3}^f)
    \frac{m_J^2}{12 m_f^2}\right|^2\,,
\end{eqnarray}
where the cancellation of the anomalous contribution has been taken
into account.

\subsection{Numerical results}

In this section we discuss some numerical results regarding the
implementation of the decaying Majoron dark matter hypothesis in our
scenario.  In Ref.~\cite{bazzocchi:2008fh} it was shown that the
experimental limit in the Majoron decay rate into photons is of the
order of $10^{-30}\ \mathrm{s}^{-1}$. It was also shown that, in a
generic seesaw model, a sizeable triplet vev plays a crucial role in
bringing the decay rate close to this experimental bound. Here we
have computed the width of the Majoron into neutrinos and photons in
our extended seesaw model which incorporates the $A_4$ flavor
symmetry, generalizing the models of Ref.~\cite{Hirsch:2007kh}.
\begin{figure}[!htb]
  \centering
  \begin{tabular}{cc}
    \includegraphics[width=0.45\linewidth]{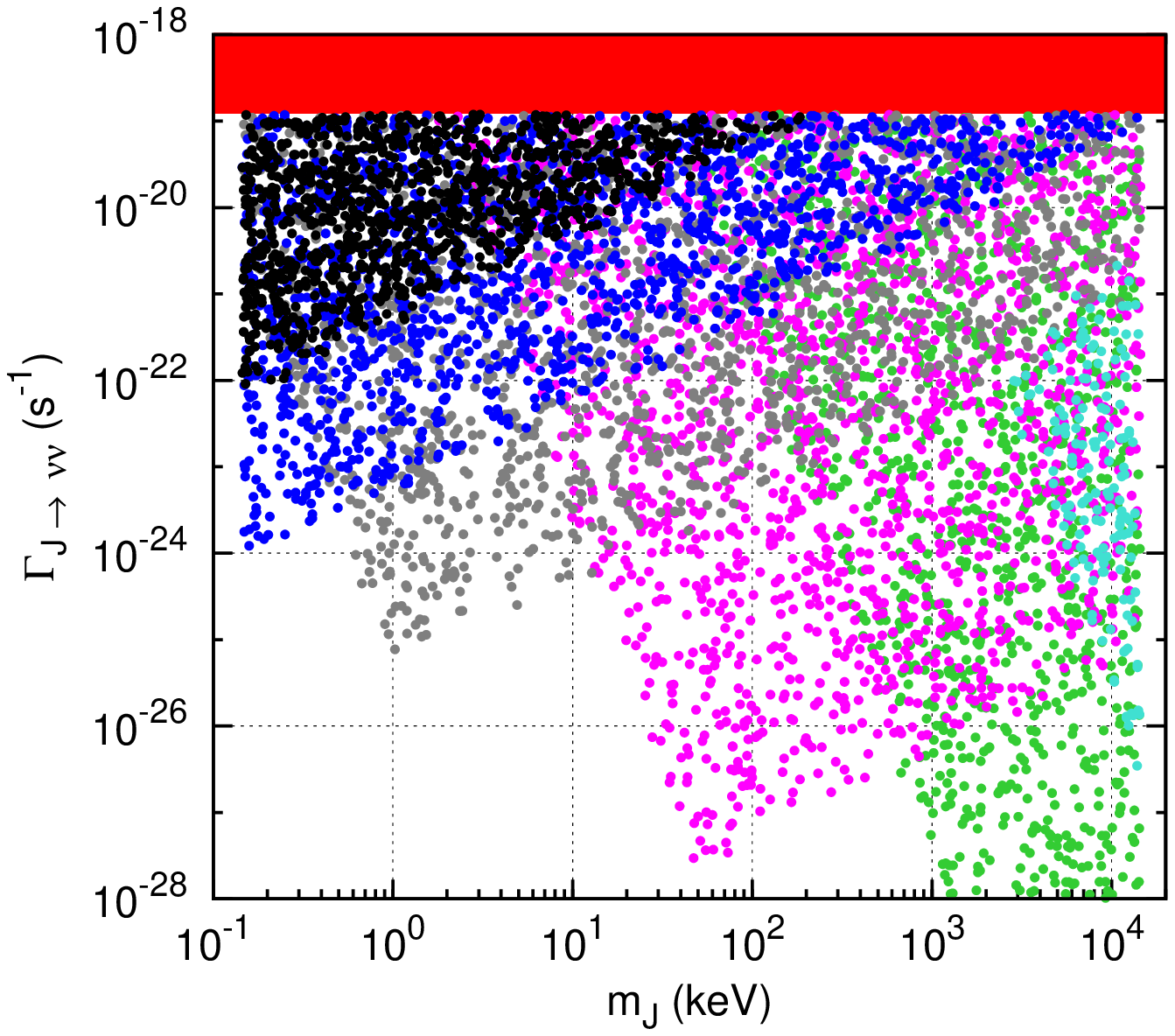}&
    \includegraphics[width=0.45\linewidth]{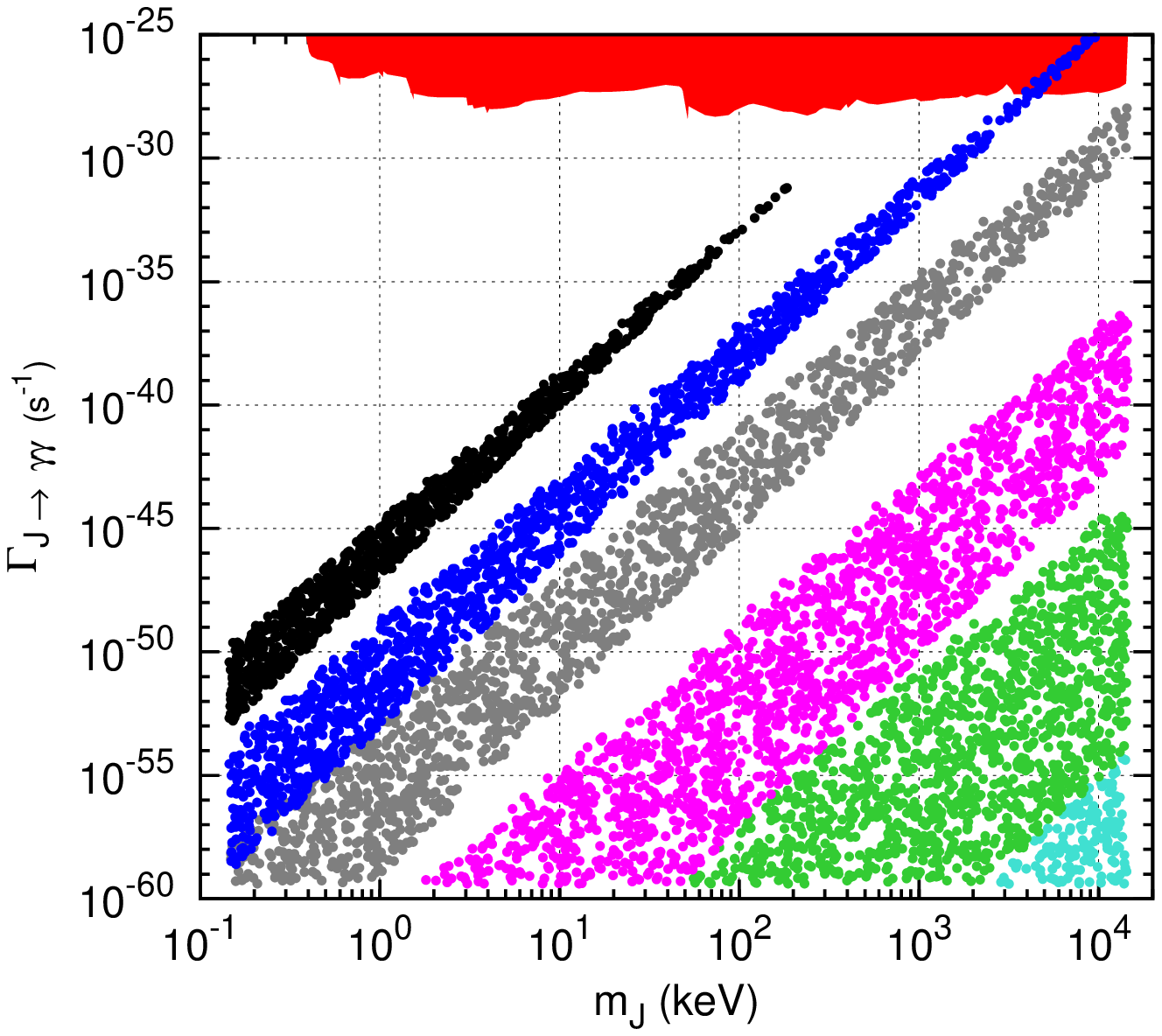}
  \end{tabular}
  \caption{Left panel: $\Gamma_{J\nu\nu}$ as function of the Majoron
    mass respecting Eq.~(\ref{eq:7}) for $u_{\Delta}=$1 eV (turquoise),
    100 eV (dark green), 10keV (magenta), 1MeV (grey), 10MeV (dark
    blue) and 100 MeV (black).  Right panel: $\Gamma_{J\gamma\gamma}$ as
    function of the Majoron mass for the same values of the triplet
    vev as in the left panel. The upper orange shaded region is the
    excluded region from X-ray observations taken from
    Ref.\cite{herder:2009im}. }
\label{fig:JNuNuJGG}
\end{figure}
The results are shown in Fig.~\ref{fig:JNuNuJGG}.  These take into
account the current neutrino oscillation data, discussed in section
\ref{sec:neutrino}.
We chose five values for the triplet vev, $u_{\Delta}=$1~eV
(turquoise), 100~eV (dark green), 10~keV (magenta), 1~MeV (grey) and
10~MeV (dark blue) and 100~MeV (black). For the right panel we
consider only points that satisfy the WMAP constraint (\ref{eq:7})
indicated by the red horizontal band on the top of the left plot.

In order to be able to probe our decaying Majoron dark matter
scenario through the mono-energetic emission line one must be close
to the present experimental limits on the photon decay channel,
discussed in Ref.~\cite{bazzocchi:2008fh} and references therein. As
mentioned, this requires the triplet vev to be sizeable, as shown on
the right panel of Fig.~\ref{fig:JNuNuJGG} for the same choices of
$u_{\Delta}$. In principle there is an additional lower bound on the
Majoron mass coming from the Tremaine-Gunn
argument~\cite{tremaine:1979we}, which, for fermionic dark matter
would be around 500~eV. Under certain assumptions this bound could
be extended to bosons, and is expected to be somewhat
weaker~\cite{Madsen:1990pe}.  The upper orange shaded region is the
excluded region from X-ray observations given in
Ref.~\cite{herder:2009im}. One should point out that, in this model,
because of the vev seesaw relation $u_\Delta u_\sigma\sim v^2$ one
cannot arbitrarily take large values for $u_{\Delta}$ to enhance
$\Gamma_{J\gamma\gamma}$ because then the singlet vev gets
correspondingly smaller values, hence reducing the lifetime of the
Majoron to values in conflict with the WMAP constraint. This
interplay between the CMB bounds and the detectability of the gamma
line is illustrated in Fig.~\ref{fig:JNuNuJGG}, where the dark-blue
points corresponding to $u_{\Delta}=10$ MeV illustrate the
experimental sensitivity to our signal.

\section{Conclusions}
\label{sec:conclusions}

We have studied the possibility that the seesaw model with
spontaneously broken ungauged lepton number may simultaneously
account for the observed neutrino masses and mixing as well as the
dark matter of the Universe.
We have presented a two-texture structure for the neutrino mass
which arises in a specific seesaw scheme implementing an $A_4$
flavor symmetry.
A predictive pattern of neutrino masses emerges from the interplay
of type-I and type-II seesaw contributions, with a lower bound on
the neutrinoless double beta decay rate, which correlates with the
deviation from maximality of the atmospheric mixing angle
$\theta_{23}$, as well as nearly maximal CP violation, correlated
with the reactor angle $\theta_{13}$.

On the other hand, assuming that associated Majoron picks up a mass
due to explicit lepton number violating effects that may arise, say,
from quantum gravity, we showed how it can constitute a viable
candidate for decaying dark matter, consistent with cosmic microwave
background lifetime constraints that follow from current WMAP
observations.
We have also shown how the Higgs boson triplet, whose existence is
required by the consistency of the model, plays a key role in
providing a test of the decaying Majoron dark matter hypothesis,
implying the existence of a mono-energetic emission line which
arises from the sub-leading one-loop-induced decay of the Majoron
into photons. We also discussed the possibility of probing its
existence in future X-ray observations such as expected in NASA's
Xenia mission~\cite{xenia}. The presence of the type-II seesaw Higgs
triplet would also have other particle physics implications, such as
lepton flavor violating decay rate enhancements due to the strong
renormalization effects of the triplet, quite sizeable in a
supersymmetric model.

\section*{Acknowledgments}

This work was supported by the Spanish MICINN under grants
FPA2008-00319/FPA and MULTIDARK CAD2009-00064 (Con-solider-Ingenio
2010 Programme), by Prometeo/2009/091, by the EU Network grants
UNILHC PITN-GA-2009-237920 and MRTN-CT-2006-035505 and by {\it
Funda\c c\~ao para a Ci\^encia e a Tecnologia} grants CFTP-FCT UNIT 777,
POCI/81919/2007 and CERN/FP/83503/2008. M.A.T. acknowledges financial 
support from CSIC under the JAE-Doc program. The work of J. N. E. is 
supported by ``Funda\c{c}\~ao para a Ci\^encia e Tecnologia" under the 
grant SFRH/BD/29642/2006.

\appendix

\section{Basic $A_4$ results}

The group $A_{4}$ consists of the even permutations of four elements
and has three one-dimensional representations and one
three-dimensional, see, e.g.~\cite{he:2006dk}. Using the usual
notation for transpositions and cyclic permutations (for instance,
(123)4 applied to $abcd$ gives $bcad$), the one-dimensional
representations are shown in Table \ref{tab:1},
\begin{table}[ht]
  \centering
\caption{Unidimensional representations for $A_4$.}
  \begin{tabular}{cccccc}\hline\hline
    &\vb{12} & \hspace{.5cm} $\mathbf{1}$&\hspace{.5cm}$\mathbf{1^{'}}$
&\hspace{.5cm}$\mathbf{1^{''}}$  \\[+2pt]  \hline
&Class 1 &  \hspace{.5cm} 1   &   \hspace{.5cm}1       &
\hspace{.5cm}  1   \\[+2pt]  \hline
&Class 2 &  \hspace{.5cm} 1   &  \hspace{.5cm} 1       &
\hspace{.5cm}  1   \\[+2pt]  \hline
&Class 3 &  \hspace{.5cm} 1   &  \hspace{.5cm} $\omega^2$  &
\hspace{.5cm}$\omega$ \\[+2pt]   \hline
&Class 4 &  \hspace{.5cm} 1   &  \hspace{.5cm} $\omega$ &
\hspace{.5cm} $\omega^2$\\[+2pt]  \hline\hline
          \end{tabular}
\label{tab:1}
\end{table}
\noindent where $\omega=e^{2\pi i/3}$ is the cubic root of unity,
and the equivalence classes are defined as

\begin{flushleft}
Class 1: e\\
Class 2: (12)(34), (13)(24), (14)(23)\\
Class 3: 1(234), 2(143), 3(142), 4(132)\\
Class 4: 1(243), 2(134), 3(124), 4(123).
\end{flushleft}

It follows immediately that
\begin{center}
$\mathbf{1'}\times\mathbf{1'}=\mathbf{1''}$,\hspace{.5cm}$\mathbf{1'}\times
\mathbf{1''}=\mathbf{1}$,\hspace{.5cm}$\mathbf{1''}\times\mathbf{1''}=
\mathbf{1'}$.
\end{center}

As for the decomposition for the tensorial product of two triplets
in $A_{4}$ one has:
\begin{equation}
 \mathbf{3} \otimes \mathbf{3} = \mathbf{1} \oplus \mathbf{1'}\oplus
 \mathbf{1''}\oplus \mathbf{3}_s\oplus\mathbf{3}_a\,,
\end{equation}
where the triplet and singlet representations are
\begin{eqnarray}
 (u\otimes v)_{\mathbf{1}}&=&u_{1}v_{1}+u_{2}v_{2}+u_{3}v_{3}\\
 (u\otimes v)_{\mathbf{1'}}&=&
u_{1}v_{1}+\omega^2 u_{2}v_{2}+\omega u_{3}v_{3}\\
 (u\otimes v)_{\mathbf{1''}}&=&
u_{1}v_{1}+\omega u_{2}v_{2}+\omega^2 u_{3}v_{3}
\\
(u\otimes v)_{\mathbf{3}_s}&=&
(u_2v_3+v_3u_2,u_3v_1+v_1u_3,u_1v_2+u_2v_1)
\\
(u\otimes v)_{\mathbf{3}_a}&=&
(u_2v_3-v_3u_2,u_3v_1-v_1u_3,u_1v_2-u_2v_1)\,.
\end{eqnarray}


\end{document}